\documentclass[prl,twocolumn,epsf,psfig]{revtex4}
\usepackage{graphicx}
\DeclareGraphicsExtensions{.pdf,.png,.gif,.jpg}

\begin{document}

\title{Three component fermion pairing in two dimensions}

\author{Theja N. De Silva}
\affiliation{Department of Physics, Applied Physics and Astronomy,
The State University of New York at Binghamton, Binghamton, New York
13902, USA.}
\begin{abstract}
We study pairing of an interacting three component Fermi gas in two
dimensions. By using a mean field theory to decouple the
interactions between different pairs of Fermi components, we study
the free energy landscapes as a function of various system
parameters including chemical potentials, binding energies, and
temperature. We find that the s-wave pairing channel is determined
by both chemical potentials and the interaction strengths between
the three available channels. We find a second order thermal phase
transition and a series of first order quantum phase transitions for
a homogenous system as we change the parameters. In particular, for
symmetric parameters, we find the simultaneous existence of three
superfluid orders as well as re-entrant quantum phase transitions as
we tune the parameters.
\end{abstract}

\maketitle

\section{I. Introduction}

Recent experimental progress achieved in ultra-cold atomic gases
allows one to set up test beds for controlled study of many body
physics. By tuning three dimensional two-body scattering length
between atoms in different hyperfine spin states of a dilute system
at low temperatures, interaction strength between atoms can be
controlled very precisely~\cite{fbexpt}. This can be done by using
magnetically tuned Feshbach resonance~\cite{fb}. Moreover,
interaction and spatial dimensionality can be effectively controlled
by applying an optical lattice. An effective two-dimensional system
can be created by applying a relatively strong one-dimensional
optical lattice to an ordinary three-dimensional system. As there
are always two body bound states exist for attractive two body
potentials in two dimensions~\cite{2dbound}, different pairs of
Fermi atoms can undergo Bose Einstein condensation and form
superfluidity at low temperatures. The two dimensional bound state
energies can be controlled by tuning either three dimensional
scattering length or the laser intensity which used to create one
dimensional lattice to accommodate 2D layers.

In this paper we study three component Fermi gases in two
dimensions. A mixture of $^6$Li atoms which has favorable
collisional properties among its lowest three hyperfine spin states
will be an ideal system to explore novel three component
superfluidity. Three component $^6$Li mixture in three dimensions
have already been trapped and manipulated experimentally~\cite{ex1,
ex2}. In Ref.~\cite{ex1}, using radio frequency spectroscopic data
and a quantum scattering model, scattering lengths and the Feshbach
resonance positions in the lowest three channels of $^6$Li atoms
have been determined. As there are three broad $s$-wave Feshbach
resonances, one can prepare the system at various interaction
strengths between each pairs. In Ref.~\cite{ex2}, collisional
stability of the lowest three channels of $^6$Li atoms has been
studied. As the spin relaxation time is large compared to the other
time scales in the experiments, experimentalists were able to
maintain a fixed spin population throughout the experiments. As a
physically accessible system, a three-component ultra-cold atomic
system can be used to study the physics of nuclear matter.
Three-component Fermi pairing is believed to occur in the interior
of neutron stars and in heavy-ion collisions~\cite{nuclear}.
Nevertheless, this system can be used to understand the competition
between quantum phases and re-entrant phase transitions.

Properties of three-component Fermi systems have been extensively
studied in recent past~\cite{paananen, paananen2, th1, th2, th3,
th4}. However, all these studies are carried out in a three
dimensional or one dimensional environments. Further, authors in all
these references except Ref.~\cite{paananen2} have restricted their
parameter space either by assuming equal interaction strengths or
equal chemical potentials or by neglecting the interaction between
some hyperfine spin components. In ref.~\cite{paananen2}, the
authors have studied the properties of a harmonically trapped three
component gas in three dimensions.

In this paper, we neglect the possibility of three body bound states
in two dimensions and consider only two body pairing states. This is
reasonable, because of the system we are considering is dilute and
the atomic interactions are short range in nature. Therefore, it is
unlikely to have many atoms interacting in the same region of space.
Further, we neglect the harmonic confinement and consider the system
as spatially homogenous. Two-component Fermi gases in two dimensions
have already been studied in theory~\cite{2d2comptheory, 2dtheja}.
The purpose of this paper is to investigate how pairing will occur
when a third spin component is added to such a two-component gas.
More precisely, we study the competition of individual components to
form Bose condensed pairs by investigating the landscapes of the
free energy as a function of various parameters which include the
temperature, chemical potentials of the hyperfine spin components,
and the interaction strengths between different pairs of fermions
components. For appropriately chosen parameters, we find that the
system undergoes a second order thermal phase transition from normal
state to a superfluid state as one lowers the temperature. At low
temperatures, we find a series of first order quantum phase
transitions as we change the chemical potentials or the interactions
between hyperfine spin components. At low temperatures, we find
simultaneous existence of three types of superfluid phases (at
symmetric parameters corresponding to different pairing channels)
and normal phases in this novel three-component Fermi system.
Interestingly, we find that the system can undergo re-entrant phase
transitions as we simultaneously tune the chemical potentials and
interactions.

The paper is organized as follows. In the following section, we
introduce the theoretical model and use a mean field approximation
to decouple the interaction terms. Then using a canonical
transformation, we diagonalize the Hamiltonian to derive the free
energy of the system. In section III, we present our results with a
discussion. Finally, our summary and conclusions are given in
section IV.

\section{II. Formalism}

\begin{figure}
\includegraphics[width=\columnwidth]{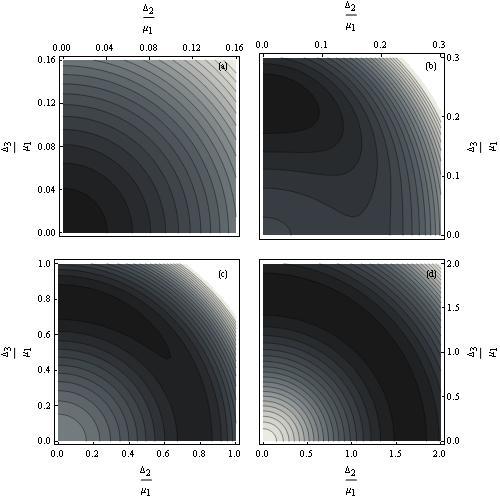}
\caption{Free energy contours showing a second order thermal phase
transition for the parameters $\mu_2=0.8 \mu_1$, $\mu_3=0.75\mu_1$,
$E_{B1}=0.1\mu_1$, $E_{B2}=\mu_1$ and $E_{B3}=0.99\mu_1$. From (a)
to (d) temperature varies as $k_BT=\mu_1/1.00$,
$k_BT=\mu_1/1.11$,$k_BT = \mu_1/1.20$, and $k_BT = \mu_1/50$. The
global minimas in figures (a) to (d) are at $(\Delta_1/\mu_1,
\Delta_2/\mu_1, \Delta_3/\mu_1) = (0, 0, 0)$, (0, 0, 0.23), (0, 0,
0.79), and (0, 0, 1.66) respectively.} \label{tpt}
\end{figure}

We consider an interacting three-component Fermi atomic gas trapped
in two dimensions.  We take the model Hamiltonian of the system as

\begin{eqnarray}\label{model}
H = \int
d^2\vec{r}\biggr\{\sum_{n}\psi^{\dagger}_{n}(r)[-\frac{\hbar^2\nabla^2_{2D}}{2m}-\mu_{n}]\psi_{n}(r)
\\ \nonumber
+\frac{1}{2}\sum_{n\neq n^\prime }U_{n
n^\prime}\psi^{\dagger}_{n}(r)\psi^{\dagger}_{n^\prime}(r)\psi_{n^\prime}(r)\psi_{n}(r)\biggr\}
\end{eqnarray}

\noindent where $r^2 = x^2+y^2$, $\nabla_{2D}$ is the 2D gradient
operator and $U_{n n^\prime}$ is the 2D interaction strength between
component $n$ and $n^\prime$. The operator $\psi^{\dagger}_{n}(r)$
creates a fermion of mass $m$ with hyperfine spin $n = 1, 2, 3$ at
position $r = (x, y)$. The chemical potential of the $n$'th
component is $\mu_n$. Notice that we have neglected the interaction
between the same components. This is reasonable as we are
considering a dilute atomic system, and the interactions are
short-range in nature, s-wave scattering channel is dominated over
the other scattering channels. By using a mean field decoupling of
the interacting terms, the mean field Hamiltonian in the momentum
space can be written as,

\begin{eqnarray}\label{mf1}
H_{MF} =
\sum_{ij}\psi^{\dagger}_{i}A_{ij}\psi_{j}+\frac{1}{2}\sum_{ij}(\psi^{\dagger}_{i}B_{ij}\psi^{\dagger}_{j}+h.c)-\sum_{i\ne
j}\frac{|\Delta_{ij}|}{U_{ij}}
\end{eqnarray}
\noindent where $i=k,n$ and $j=-k,n$. Here we defined the superfluid
order parameters $U_{ij}\langle \psi_i \psi_j \rangle=\Delta_{ij}$
and two matrices $A$ and $B$,

\begin{eqnarray}\label{a}
A= \left(
                     \begin{array}{ccc}
                       \epsilon_1 & 0 & 0\\
                       0 & \epsilon_2 & 0 \\
                       0 & 0 & \epsilon_3 \\
                     \end{array}
\right)
\end{eqnarray}

\begin{eqnarray}\label{a}
B= \left(
                     \begin{array}{ccc}
                       0 & \Delta_{3} & -\Delta_{2}\\
                       -\Delta_{3} & 0 & \Delta_{1} \\
                       \Delta_{2} & -\Delta_{1} & 0 \\
                     \end{array}
\right)
\end{eqnarray}

\noindent where $\epsilon_n=\hbar^2k^2/(2m)-\mu_n$ and
$\Delta_{ij}=\epsilon_{ijk}\Delta_k$. As the mean field Hamiltonian
is quadratic in Fermi operators, it can be diagonalized with a
canonical transformation to get

\begin{eqnarray}\label{mf2}
H_{MF} = \sum_{n,
k}\Lambda_n\eta^{\dagger}_{n}\eta_{n}+\frac{1}{2}\sum_{n,
k}(\epsilon_n-\Lambda_n)-\sum_{n}\frac{|\Delta_{n}|}{U_{n}}
\end{eqnarray}

\noindent where we use the notation $U_{ij}=|\epsilon_{ijk}|U_k$.
The Fermi operators $\eta_n$ represent the quasi particles in the
system with $n=1, 2, 3$. The quasi particle energies $\Lambda_n =
\sqrt{\lambda_n}$ are given by $\lambda_n = 2\sqrt{-Q}
\cos\{[\theta+(n-1)2\pi]/3\}-A_k/3$. The parameter $\theta = \arccos
[R/\sqrt{-Q^3}]$ with $Q=(3B_k-A_k^2)/9$ and
$R=(9A_kB_k-27C_k-2A_k^3)/54$~\cite{eigenvalue}. Here we defined
$A_k = -\sum_n\epsilon_n^2-2\sum\Delta_n^2$, $B_k =
\sum_n\Delta_n^4+2\sum_n\Delta_n^2\epsilon_n^2 + \sum_{n\ne
m}\Delta_n^2\Delta_m^2 + 1/2\sum_{n\ne m}\epsilon_n^2\epsilon_m^2 +
\sum_{n\ne m\ne l}\Delta_n^2\epsilon_m\epsilon_l$ and $C_k =
-(\sum_n\Delta_n^2\epsilon_n+\epsilon_1\epsilon_2\epsilon_3)^2$. The
grand potential of the system $\Omega=-1/(\beta)\ln[Z_G]$ with
$Z_G=tr\{e^{[-\beta H_{MH}]}\}$ is then given by

\begin{eqnarray}\label{mf2}
\Omega = -1/(\beta)\sum_{n, k}[\ln(1+e^{-\beta\Lambda_n})]
\\ \nonumber +\frac{1}{2}\sum_{n, k}(\epsilon_n-\Lambda_n)-\sum_{n}\frac{|\Delta_{n}|}{U_{n}}
\end{eqnarray}

\noindent where $\beta=1/(k_BT)$ is the inverse temperature and
$k_B$ is the Boltzmann constant. As the short range nature of the
interaction, the grand potential is diverging so that regularization
must be done in standards way by writing $-|\Delta_n|^2/U_n =\sum_k
|\Delta_n|^2/(\hbar^2k^2/m+E_{Bn})$. Here
$E_{Bij}=|\epsilon_{ijk}|E_{Bk}$ is the binding energy between two
hyperfine spin components $i$ and $j$. Notice that we use the same
notation for $E_{Bn}$ as we used for
$\Delta_{ij}=\epsilon_{ijk}\Delta_k$ and
$U_{ij}=|\epsilon_{ijk}|U_k$. For two dimensions, converting the
$\sum_k$ into integral $\int d^2k/(2\pi)^2$ and then by changing the
variable by $k^2 = z$, the grand potential can be converted into an
one dimensional integral. We numerically perform this integral and
numerically minimize the grand potential for the seven parameter
space (three chemical potentials, three binding energies and the
temperature).

\section{III. Results and discussion}

\begin{figure}
\includegraphics[width=\columnwidth]{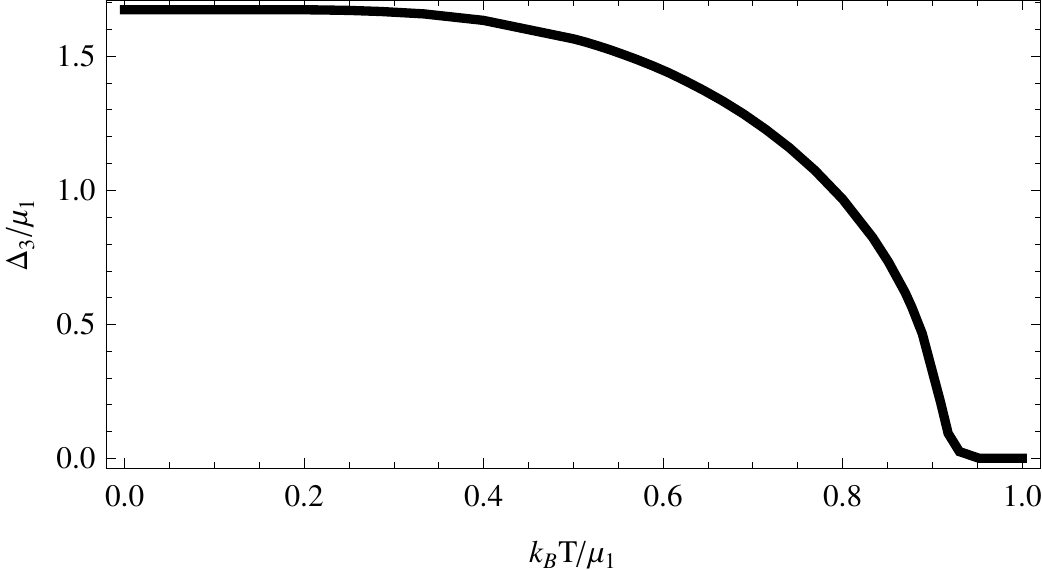}
\caption{Superfluid order parameters $\Delta_3$ as a function of
temperature. We fixed the binding energies $E_{B1} =0.1 \mu_1$,
$E_{B2} = 0.5 \mu_1$, $E_{B3} = \mu_1$ and chemical potentials
$\mu_2 = 0.8 \mu_1$ and $\mu_3 = 0.75 \mu_1$.} \label{qd1}
\end{figure}

In Fig.~\ref{tpt}, we plot the free energy landscapes for different
temperatures at a selected set of parameters. We choose the binding
energy in the $2 - 3$ channel to be small $(E_{B1} = 0.1 \mu_1)$ so
that pairing is not possible in this channel. As a result, we find a
second order thermal phase transition as we lower the temperature.
As can be seen, at high temperature [Fig.~\ref{tpt}-(a)], free
energy is minimum when both paring order parameters ($\Delta_2$ and
$\Delta_3$) in channels $1 - 3$ and $1 - 2$ are zero. As we lower
the temperature, channel $1 - 2$ undergoes pairing and form Bose
condensation. This is because the binding energy and average
chemical potential in this channel is larger than those of channel
$1 - 3$ and $2 -3$. Further lowering the temperature results more
atom pairing and condensation in channel $1 - 2$ giving larger
superfluid order parameter $\Delta_3$. In principle, it is possible
to have a sequence of second order thermal and first order quantum
phase transitions at three different critical temperatures, if one
change the interactions or the chemical potentials together with
temperature. The reason for this sequence of phase transition is
that there are many ways of pairing when various favorable channels
are available.

In Fig.~\ref{qd1}, we plot the temperature dependence of the
superfluid order parameter $(\Delta_3)$ in channel $1 - 2$ for
chosen values of parameters. We choose the parameters such that
pairing is possible only in channel $1 - 2$ so that a single minimum
is available in the free energy. As can be seen, the superfluid
order parameter continuously increases as one lower the temperature,
showing a second order thermal phase transition.

\begin{figure}
\includegraphics[width=\columnwidth]{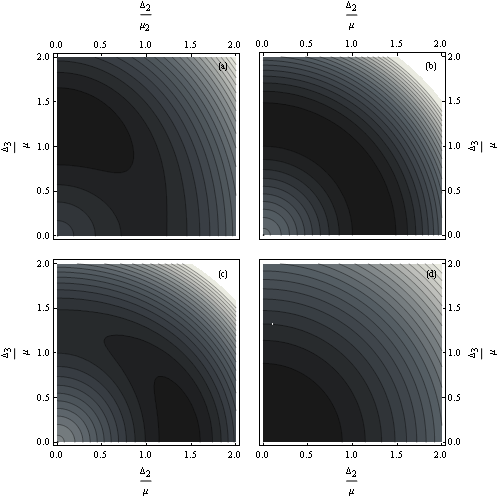}
\caption{Low temperature free energy contours showing first order
quantum phase transitions for the parameters $\mu_1= \mu_2 = \mu_3 =
\mu$,
 $E_{B1}=0.1\mu$, $E_{B3}=0.6\mu$ and $k_BT = \mu/50$. From panel (a) to (c)
 $E_{B2}$ varies as $0.4\mu$, $0.6\mu$, $0.7\mu$ respectively. In panel (a) and (c) the minimas are given at $(\Delta_1/\mu, \Delta_2/\mu, \Delta_3/\mu) = (0, 0, 1.25)$
 and $(0, 1.37, 0)$ respectively. In panel (b), free energy gives many stationary points where the global minimum is a quarter of a circle. In panel (d), we use $E_{B2} =E_{B3} = 0.05\mu$
and the minimum is given at (0.46, 0, 0).} \label{qpt1}
\end{figure}

By varying the average chemical potentials and binding energies of
the pairing channels at low temperatures, one can control the first
order quantum phase transitions from one superfluid phase to
another. As a demonstration, we plot the free energy landscapes in
Fig.~\ref{qpt1} for a selected set of parameters. Again, we chose
the parameters such that the pairing in channel $2 - 3$ is very weak
and the corresponding superfluid order parameter $(\Delta_1)$ is
zero. When the binding energy in channel $1 - 2$ is larger than that
of the channel $1 -3$, the free energy minimum is at a non zero
value of $\Delta_3$ but zero value of $\Delta_2$ at the same
chemical potentials. However, when the binding energies are equal at
equal average chemical potentials, both superfluid order parameters
are non zero and free energy gives many stable stationary points as
seen in Fig.~\ref{qpt1}-(b) (now the global minimum is not a point,
but a quarter of a circle). The reason for this line of global
minimum is the symmetry of the parameter space. By increasing the
binding energy in channel $1 -3$ over the channel $1 -2$, the
minimum free energy pass to the non zero $\Delta_2$ but zero
$\Delta_3$. Similar first order quantum phase transitions can be
seen by controlling the average chemical potentials at fixed and
equal binding energies. More generally, by controlling the chemical
potentials and binding energies, one can observe not only a series
of quantum phase transition, but a phase with multi-component
superfluid order (simultaneous existence of three superfluid order
parameters). Notice that one can have a re-entrant quantum phase
transition by changing both chemical potentials and binding energies
simultaneously.

\begin{figure}
\includegraphics[width=\columnwidth]{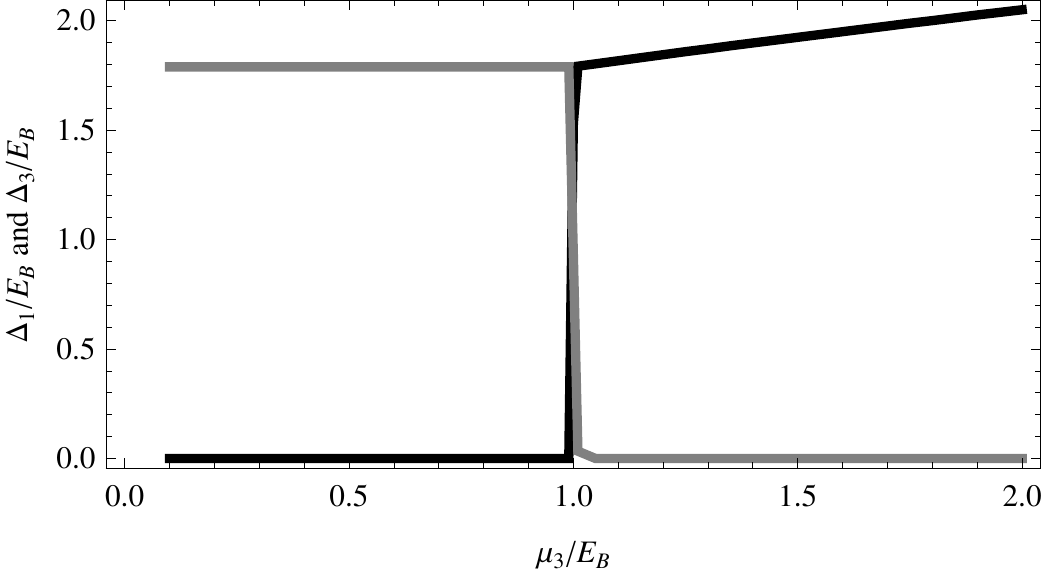}
\caption{Superfluid order parameters $\Delta_1/E_B$ (black line) and
$\Delta_3/E_B$ (gray line) as a function of $\mu_3/E_B$. We fixed
the binding energies $E_{B1} =E_{B2} = E_{B3} =E_{B}$ and the
temperature $k_BT=E_B/50$. The chemical potentials are $\mu_1 = E_B$
and $\mu_2 = 1.2 E_B$. In the entire range of $\mu_3$, $\Delta_2$ is
zero.} \label{dsb}
\end{figure}

In Fig.~\ref{dsb}, we plot superfluid order parameters as a function
of the chemical potential of the third component. The binding
energies are fixed to be the same for all three channels. As can be
seen in figure, fermions pairing occurs in channel $1 -2$ at smaller
$\mu_3$. This is because the average chemical potential in this
channel is the largest for $\mu_3 < E_B$. Further, as the average
chemical potential is constant, superfluid order parameter
$\Delta_3$ is constant. For $\mu_3 > E_B$, average chemical
potential in channel $2 - 3$ is the largest and increasing with
increasing $\mu_3$. As a result, superfluid order parameter
$\Delta_1$ increases with $\mu_3$. In the entire range of $\mu_3$,
average chemical potential in channel $1 - 3$ is smaller than that
of the other channels so that the pairing in this channel is not
favorable. As seen in Fig.~\ref{dsb}, the superfluid order
parameters have large discontinuity which represents a sharp first
order quantum phase transition.

\begin{figure}
\includegraphics[width=\columnwidth]{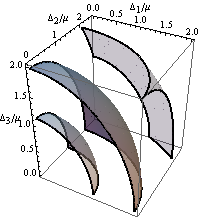}
\caption{Superfluid order parameters for symmetric parameters. We
use the same chemical potentials for the three species
$\mu_1=\mu_2=\mu_3=\mu$ and same interaction strengths for the three
channels $E_{B1}=E_{B2}=E_{B3}= E_B$. The three surfaces
($\Delta_1^2+\Delta_2^2+\Delta^2_3 =$ constant) shown in the order
parameter space are for $E_B= 0.5\mu$, $1.2\mu$, and $2.0\mu$. We
fix the temperature to be $k_BT=\mu/50$.} \label{con3d}
\end{figure}

At the same chemical potentials and the same interaction strengths
of the channels, free energy gives many stable stationary points at
which the condition $\Delta_1^2+\Delta_2^2+\Delta^2_3 = C$ is
satisfied. The constant $C$ depends on both the chemical potentials
and the interaction strengths (binding energies). As shown in
Fig.~\ref{con3d}, when the free energy has a minimum, the order
parameters represent a surface in order parameter space.

Within our mean field description, we were able to handle only
two-body correlations. One needs to go beyond mean field theory to
understand the role of three-body correlations in a three-component
system. If three atoms can overlap in the same region of space, then
the three-body correlations can play a role giving Thomas
effect~\cite{thomas effect} and Efimov effect~\cite{efimov}. Thomas
effect is the collapse of a three body system due to the overlap of
atoms. Atom loss in a trap is undoubtedly related to the Thomas
effect. In a three component atomic system, collision between a
condensed pair and a third species atom can support the Thomas
effect. Efimov effect is the accumulation of three-body bound states
at strongly interacting limit. These effects are forbidden in two
component gases. For sufficiently low densities, these effects are
forbidden even in three component systems so that our results are
applicable to dilute ultra-cold atomic gases. We discussed the pure
2D limit in this paper, however one can generalize the theory to
include the weak atom tunneling between layers as done in
Ref.~\cite{2dtheja} for two component gases.

Our results in two dimensions look qualitatively similar to the ones
obtained in three dimensions in Ref.~\cite{paananen}. However, we
find that the superfluidity is more sensitive to the parameters in
two dimensions than three dimensional systems. As we have seen
above, superfluidity is very sensitive to both chemical potentials
and the interactions between different pairs. In current
experimental setups, Feshbach resonance allows one to control the
interactions to different values by tuning the scattering lengths.
However, the scattering lengths between different pairs of fermions
cannot be controlled independently. Therefore, chemical potential is
the suitable parameter to drive the quantum phase transitions.
Typically, chemical potentials can be controlled by changing the
atomic population in both two dimensions and three dimensions.
However, in order to change the chemical potentials this way, one
has to start the experiment all over with a different atomic sample.
The advantage of using quasi two dimensional system is that one can
change the effective chemical potential by controlling the tunneling
between layers. This tunneling can be controlled by the laser
intensity of the optical lattice. As shown in Ref.~\cite{2dtheja2},
the first order quantum phase transitions in two dimensional systems
can easily be controlled by the laser intensity.

\section{IV. Summary and Conclusions}

We studied Fermi superfluidity of an interacting three component
system in two dimensions. We used a mean field theory to investigate
the behavior of free energy as a function of chemical potentials,
binding energies, and the temperature. Depending on the chemical
potentials and binding energies, we find a second order thermal
phase transition as we lower the temperature. At low temperatures,
by controlling the parameters, on can have a series of first order
quantum phase transitions and re-entrant phase transitions. These
series of phase transitions are associated with pairing between
different hyperfine spin components of fermions.

The possible pairing is determined by both average chemicals
potentials and the interaction strengths of the paring channels. The
channel which has largest paring strength forms the superfluid,
while the unpaired component form a Fermi sea. At low temperature,
first order quantum phase transition can be induced by increasing
the average chemical potential or the interaction strength of one
channel over the other. If the pairing strengths are equal in all
three channels, then it is possible to have three superfluid phases
simultaneously. As we have not considered the interaction between
condensed pairs, we do not expect the phase separation of superfluid
phases in spatially homogenous environments~\cite{th1}. However in
trapped systems, it has been shown that the phase separation of
superfluid phases is possible in three dimensions~\cite{paananen2}.

We speculate that the simultaneous existence of three types of
superfluid phases in trapped systems can be detected by standard
experimental methods. For example, superfluidity can be demonstrated
by the creation of vortices~\cite{vortices} and then distinguished
them by probes coupling to each atom types. Alternatively, one can
measure the energy gap using radio frequency spectroscopy~\cite{rf}
or measure condensate fraction using the pair projection
method~\cite{pair}.

\end{document}